\documentclass[a4paper,fleqn,usenatbib,useAMS]{mnras}
\usepackage[T1]{fontenc}
\usepackage{ae,aecompl} 

\usepackage{graphicx}
\usepackage{amsmath}
\usepackage{amssymb}

\usepackage{upgreek}
\usepackage{url}

\newcommand{\teff}{\mbox{$T_\mathrm{eff}$}}
\newcommand{\logg}{\mbox{$\log g$}}

\title[Pulsation versus metallicism in Am stars]{Pulsation versus metallicism in
Am stars as revealed by LAMOST and WASP}

\author[B.~Smalley et al.]
{B.~Smalley$^{1,}$\thanks{Email: b.smalley@keele.ac.uk},
V.~Antoci$^{2}$,
D.~L.~Holdsworth$^{3}$,
D.~W.~Kurtz$^{3}$,
S.~J.~Murphy$^{4,2}$,  \newauthor
P.~De~Cat$^{5}$,
D.~R.~Anderson$^{1}$,
G.~Catanzaro$^{6}$,
A.~Collier~Cameron$^{7}$,
C.~Hellier$^{1}$, \newauthor
P.~F.~L.~Maxted$^{1}$,
A.~J.~Norton$^{8}$,
D.~Pollacco$^{9}$,
V.~Ripepi$^{10}$,
R.~G.~West$^{9}$, \newauthor
P.~J.~Wheatley$^{9}$
\\
$^{1}$Astrophysics Group, Lennard-Jones Laboratories, Keele University, Staffordshire ST5 5BG, United Kingdom\\
$^{2}$Stellar Astrophysics Centre, Department of Physics and Astronomy, Aarhus University, 8000 Aarhus C, Denmark\\
$^{3}$Jeremiah Horrocks Institute, University of Central Lancashire, Preston PR1 2HE, United Kingdom\\
$^{4}$Sydney Institute for Astronomy (SIfA), School of Physics, The University
of Sydney, NSW 2006, Australia\\
$^{5}$Royal Observatory of Belgium, Av Circulaire 3-Ringlaan 3, B-1180 Brussels, Belgium\\
$^{6}$INAF-Osservatorio Astrofisico di Catania, Via S.Sofia 78, I-95123
Catania, Italy\\
$^{7}$SUPA, School of Physics \& Astronomy, University of St. Andrews, North
Haugh, Fife, KY16 9SS, UK\\
$^{8}$Department of Physical Sciences, The Open University, Walton Hall, Milton
Keynes, MK7 6AA, UK\\
$^{9}$Department of Physics, University of Warwick, Coventry, CV4 7AL, UK\\
$^{10}$INAF-Osservatorio Astronomico di Capodimonte, Via Moiariello 16, I-80131
Napoli, Italy\\
}

\date{Accepted 2016 November 06. Received 2016 November 06; in original form 2016 September 05}

\begin{document}

\maketitle

\begin{abstract} We present the results of a study of a large sample of A and Am
stars with spectral types from LAMOST and light curves from WASP. We find that,
unlike normal A stars, $\delta$\,Sct pulsations in Am stars are mostly confined
to the effective temperature range $6900 < {\teff} < 7600$\,K. We find evidence
that the incidence of pulsations in Am stars decreases with increasing
metallicism (degree of chemical peculiarity). The maximum amplitude of the
pulsations in Am stars does not appear to vary significantly with metallicism.
The amplitude distributions of the principal pulsation frequencies for both A
and Am stars appear very similar and agree with results obtained from {\it
Kepler} photometry. We present evidence that suggests turbulent pressure is the main
driving mechanism in pulsating Am stars, rather than the $\kappa$-mechanism,
which is expected to be suppressed by gravitational settling in these stars.
\end{abstract}


\section{Introduction}

The metallic-lined (Am) stars are a subset of the A-type stars that exhibit weak
calcium K lines and enhanced iron-group spectral lines compared to their hydrogen-line
spectral type \citep{1940ApJ....92..256T,1948ApJ...107..107R}. They are further
divided into {\em classical} and {\em marginal} Am stars. The spectral
types obtained from the calcium K line ($k$) and iron-group elements ($m$) in
classical Am stars differ by at least 5 spectral sub-types, whereas in the
marginal Am stars the difference is less. There are also the `hot' Am stars with
spectral types around A0--A3, such as Sirius \citep{1964ZA.....60..115K}. As a
group the Am stars rotate relatively slowly, which is thought to be a requisite
condition for their chemical peculiarities due to radiative diffusion
\citep{1970ApJ...160..641M}. Many of the Am stars are in relatively short period
binary systems
\citep{1967mrs..conf..173A,2007MNRAS.380.1064C,2014A&A...564A..69S} and tidal
synchronisation is thought to be responsible for the observed low rotation rates
in most Am stars \citep{1973ApJ...182..809A,1983aspp.book.....W}. The results of
the binarity studies indicate that there are some apparently single Am stars
that were presumably born with slow rotation
\citep{2014A&A...564A..69S,2015MNRAS.448.1378B}.

For many years it was thought that classical Am stars did not pulsate
\citep{1970ApJ...162..597B,1976ApJS...32..651K} due to the gravitational
settling of helium from the He\,\textsc{ii} ionisation zone where the
$\kappa$-mechanism drives the pulsation of $\delta$\,Sct stars (e.g.
\citealt{2010aste.book.....A}). However, over time evidence emerged
that some Am stars pulsate
\citep[e.g.][]{1989MNRAS.238.1077K,2005AJ....129.2026H}. Recent studies using
{\it Kepler} and WASP photometry \citep{2011MNRAS.414..792B,2011A&A...535A...3S}
have found that a significant fraction of Am stars do pulsate, but with a
suspicion that they may do so at smaller amplitudes than the normal abundance
$\delta$\,Sct stars.

The Large Sky Area Multi-Object Fiber Spectroscopic Telescope
\citep[LAMOST;][]{2012RAA....12..723Z} survey is providing a large catalogue of
low-resolution stellar spectra. These are being automatically fitted to provide
an homogeneous determination of stellar parameters for A, F, G and K stars:
effective temperature ($T_{\rm eff}$) and surface gravity ($\log g$)
\citep{2011RAA....11..924W}. \citet{2015MNRAS.449.1401H} presented a list of
candidate Am stars using LAMOST data release 1 (DR1) spectra
\citep{2015RAA....15.1095L}. From the 38\,485 A and early-F stars in their
LAMOST sample, they identified 3537 Am candidates and gave $k$ and
$m$ spectral types. 

\begin{figure}
\includegraphics[width=\columnwidth]{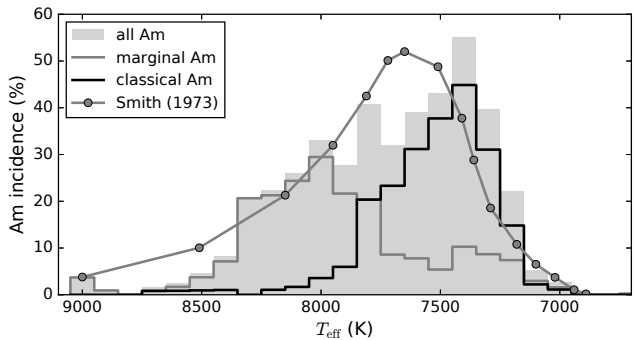}
\caption{The percentage incidence of Am stars as a function of {\teff}.
The distribution given by \citet{1973ApJS...25..277S} has been transformed from
his $b-y$ to {\teff} using the $uvby$ grids of
\citet{1997A&A...328..349S}. }
\label{teff-Am-dist}
\end{figure}

The $\Delta$ index presented by \citet{2015MNRAS.449.1401H} is defined as the
numerical difference in the $k$ and $m$ spectral types. This value is used as
the {\it metallicism} index in this study to separate classical Am stars from
the marginal ones. Values of $\Delta \ge 5$ are indicative of classical Am
stars, while we use $1 \le \Delta < 5$ for marginal Am stars.
Fig.~\ref{teff-Am-dist} shows that the incidence of Am stars in this study as a
function of {\teff} is consistent with the earlier study of
\citet{1973ApJS...25..277S} and that the marginal and classical distributions
are somewhat bimodal, with classical Am stars prevalent at lower temperature. As
also seen in \citet{1973ApJS...25..277S}, the incidence of Am stars drops
dramatically for stars cooler than ${\teff} \simeq 7000$\,K. 

In the present work we present the results of a search of the A and early-F
stars in the LAMOST DR1, as used by \citet{2015MNRAS.449.1401H} in their search
for Am stars, which have sufficient Wide Area Search for Planets
\citep[WASP,][]{2006PASP..118.1407P} photometry to obtain the pulsation
characteristics of the sample. The sample also allows for the investigation of
pulsation incidence as a function of metallicism. In the context of this study
the A and early-F stars not listed as Am candidates will be referred to as
``other A stars'' rather than ``normal'' A and F stars, since the sample
contains a contribution from other chemically peculiar stars, e.g. Ap and
$\lambda$ Boo stars \citep[e.g.][]{1983aspp.book.....W}.

\section{Sample Selection}

The WASP project has been surveying the sky for transiting extrasolar planets
using two robotic telescopes, one at the
Observatorio del Roque de los Muchachos on the island of La Palma in the Canary
Islands, and the other at the Sutherland Station, South African Astronomical
Observatory. Both telescopes consist of an array of eight 200-mm, f/1.8 Canon
telephoto lenses and Andor CCDs, giving a field of view of $7.8\degr \times
7.8\degr$ and pixel size of around 14\,arcsec. The observing strategy is such
that each field is observed approximately every 10\,min, on each observable
night. WASP provides good quality photometry with a precision
better than 1~per~cent per observation in the approximate magnitude range $9 \le V
\le 12$.

The WASP data reduction pipeline is described in detail in
\citet{2006PASP..118.1407P}. The aperture-extracted photometry from each camera
on each night is corrected for primary and secondary extinction, instrumental
colour response and system zero-point relative to a network of local secondary
standards. The resultant pseudo-$V$ magnitudes are comparable to Tycho $V$
magnitudes. Additional systematic errors affecting all the stars are identified
and removed using the {\sc SysRem} algorithm of \citet{2005MNRAS.356.1466T}.

For this study, the WASP archive was searched for light curves of stars with
coordinates coincident with those of the LAMOST spectra to within 10\,arcsec and
with the requirement that at least 1000 photometric data points were available.
Following the procedures of \citet{2011A&A...535A...3S} and
\citet{2014MNRAS.439.2078H}, amplitude spectra were calculated using the fast
computation of the Lomb periodogram method of \citet{1989ApJ...338..277P} as
implemented in the Numerical Recipes {\sc fasper} routine
\citep{1992nrfa.book.....P}. The periodograms were calculated covering the
frequency range 0 to 150 d$^{-1}$ for 30\,282 stars in the LAMOST DR1 sample.
Each periodogram was automatically searched for amplitude peaks with a
signal-to-noise ratio (SNR) greater than 4 relative to the median background
amplitude noise level and a false alarm probability (FAP) less than 0.1. For a
peak to be deemed real it must be present in more than one season of WASP
photometry and have the same frequency to within the Rayleigh criterion ($\sim
0.01$\,d$^{-1}$).

The diurnal gaps between successive observing nights introduce considerable
sampling aliases in the amplitude spectra of most WASP light curves. Peaks which
had a frequency within a Rayleigh criterion of the sidereal day frequency were
rejected. Alias peaks due to frequencies $n$ times the sidereal day frequency
were also rejected, but with the Rayleigh criterion reduced by $\sqrt{n}$. This
$\sqrt{n}$ reduction was obtained empirically, based on an examination of all
the peaks found in the sample, and models the diminishing frequency width of
an alias peak above the background noise level as $n$ increases. This
considerably reduced the number of false-positive frequencies selected, but did
not completely eliminate them. In addition, the same empirical relationship was
used to remove the day aliasing of other found frequencies.

As the WASP pixels are relatively large, blending can be an issue, especially
for fainter targets. By applying a requirement that no target shall be blended
by more than 20~per~cent within a 64\,arcsec radius, the number of stars in the
sample was reduced to 10\,525, with 864 (9.3~per~cent) being from the
\citet{2015MNRAS.449.1401H} Am star list. The blending calculation was performed
using the $R$ magnitude photometry given in the Naval Observatory Merged
Astrometric Dataset (NOMAD) catalog \citep{2004AAS...205.4815Z}.

Of the 10\,525 stars in our sample, around 1500 were identified as pulsation
candidates. These were subjected to a separate more intensive analysis,
involving a non-linear least-squares sinusoidal fit to the individual light curves, in
order to obtain the frequencies and amplitudes of the signals present. In this work
the term amplitude refers to the semi-amplitude and not the peak-to-peak light variation.
Up to a maximum of 5 frequency--amplitude pairs were selected; fewer if the FAP rose
above 0.1. Finally, the stars were visually inspected with {\sc period04}
\citep{2005CoAst.146...53L} to confirm their variability characteristics.

\begin{figure}
\includegraphics[width=\columnwidth]{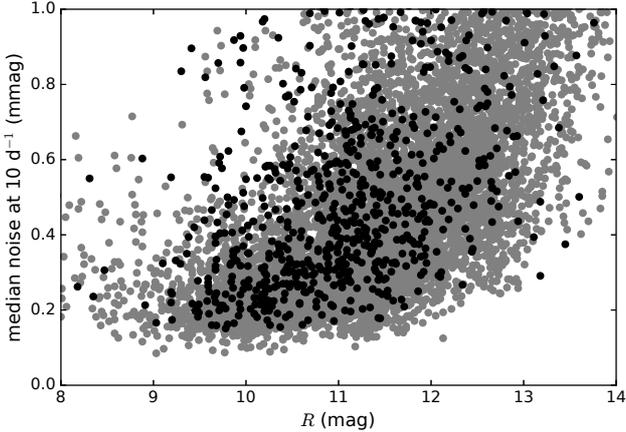}
\caption{The WASP median periodogram background noise at 10\,d$^{-1}$ against
$R$ magnitude. The black dots are those stars found to pulsate, while the grey dots are those
which were not found to pulsate.}
\label{noise}
\end{figure}

The detection threshold for WASP data was estimated by
\citet{2011A&A...535A...3S} to be around 1\,mmag. A more in-depth study by
\citet{2015PhDT.......227H} confirmed this and provided limits as a function of
stellar magnitude. Fig.~\ref{noise} shows the WASP median periodogram background
noise at 10\,d$^{-1}$, where the noise level is taken as the median value of the
periodogram within a bin of width 1\,d$^{-1}$. Since our selection criterion for
pulsations is that they must be present in at least two seasons of data, the
seasons with the lowest noise levels are used and the noise values shown are the
higher of these two. The requirement of SNR~$\ge4$ indicates that the detection
limit is at best 0.6\,mmag for brighter stars, but varies significantly from
star to star and falls off at the fainter end to above 2\,mmag. Given that the
LAMOST sample peaks around $R=11$, a typical detection limit of no better than
1\,mmag is expected, with a considerable tail.

\section{Results}

\begin{table*}
\caption{The results of the search for variability in WASP light curves. In
column 2, $n$ is number of frequencies ($f$)
and peak amplitudes ($a$) listed (up to 5). Column 5 gives the variable type inferred from the
frequency analysis (see text for details). Only the first 10 lines are shown
here. The full 1193-line table is available online.}
\scriptsize
\begin{tabular}{rrrrrrrrrrrrrrr} \\ \hline
WASP ID&$n$&$f_1$&$a_1$&variable &$f_2$&$a_2$&$f_3$&$a_3$&$f_4$&$a_4$&$f_5$&$a_5$ \\
 & & [d$^{-1}$] & [mmag] & type & [d$^{-1}$] & [mmag]& [d$^{-1}$] & [mmag]& [d$^{-1}$] & [mmag]& [d$^{-1}$]& [mmag] \\ \hline
1SWASPJ000051.83+330532.9&1&35.4725&1.22&$\delta$~Sct&&&&&&&&\\
1SWASPJ000246.71+160538.3&5&14.9049&2.83&$\delta$~Sct&17.8839&2.24&20.6246&1.97&14.7178&1.52&26.2976&1.32\\
1SWASPJ000331.13+123225.9&5&1.2281&9.12&$\gamma$~Dor&2.1342&7.42&1.1487&6.93&1.1140&5.13&1.2637&5.00\\
1SWASPJ000342.56+160511.0&2&3.8085&30.06&EB&1.9072&3.19&&&&&\\
1SWASPJ000444.67+304222.1&4&0.5579&5.33&$\gamma$~Dor&0.5666&4.51&0.6727&4.08&0.3887&3.64&&\\
1SWASPJ000534.61+292745.0&1&15.8608&2.02&$\delta$~Sct&&&&&&&&\\
1SWASPJ000613.54+362658.2&1&4.8405&65.22&$\delta$~Sct&&&&&&&&\\
1SWASPJ000811.49+322128.7&4&0.4707&12.89&$\gamma$~Dor&0.6579&9.96&0.7926&9.01&0.6324&5.45&&\\
1SWASPJ000924.89+031249.6&4&2.0414&10.18&$\gamma$~Dor&2.0879&5.90&1.9436&3.50&2.3354&2.53&&\\
1SWASPJ001430.18+365226.0&2&4.6587&103.51&$\delta$~Sct&9.3189&17.68&&&&&&\\
\hline
\end{tabular}
\label{Table_Results}
\end{table*}

The results from the frequency analysis are presented in Table~\ref{Table_Results}.
Several eclipsing binary systems were identified and labelled using the generic
`EB' label. The resistant mean method used to clean the WASP data to find
pulsations has the side effect of removing deep eclipses from the light curves
\citep{2014MNRAS.439.2078H}. Stars were also labelled as being binary systems
when a sub-harmonic of the principal frequency was found. Stars with
low-frequency ($\la$ 2\,d$^{-1}$) signals were investigated using {\sc period04}
to ascertain whether the variations could be due to other physical causes,
instead of ellipsoidal variations or pulsations. These were labelled as `Misc'
in Table~\ref{Table_Results}. Often the classification of the variability is
inconclusive in WASP data due to the noise level.

\begin{figure}
\includegraphics[width=\columnwidth]{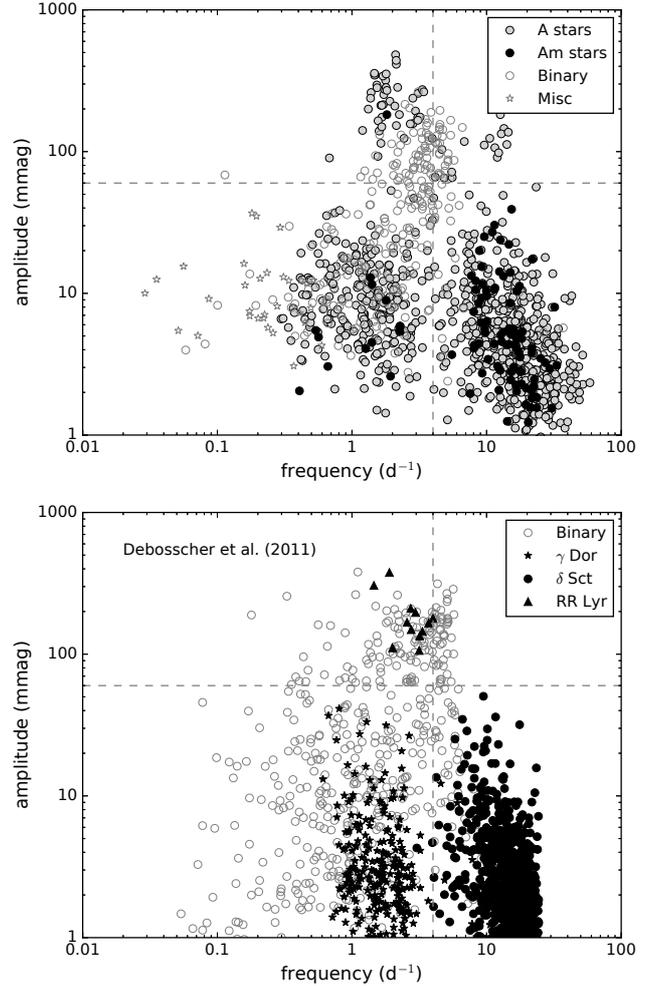}
\caption{Frequency--amplitude diagram for the stars found to exhibit variability
in the WASP photometry (top panel). The dashed lines indicate the four regions discussed in the text.
The lower panel shows the {\it Kepler} results from \citet{2011A&A...529A..89D} for stars in
the in range $6000<{\teff}<10000$\,K for comparison.}
\label{freq-amp}
\end{figure}

The frequency--amplitude distribution for the variable stars is shown in
Fig.~\ref{freq-amp}, with the {\it Kepler} results from
\citet{2011A&A...529A..89D} for stars with temperatures in the range
$6000 < {\teff} < 10000$\,K shown for comparison. A frequency of 5\,d$^{-1}$ for
the division between $\delta$~Sct and $\gamma$~Dor stars is sometimes used.
However, inspection of the WASP results suggests that 4\,d$^{-1}$ might be a
better choice, as this corresponds to the minimum in the number of pulsators
between the frequency domains where the two types mostly occur. Theory predicts
that $\gamma$~Dor stars stars rotating with 50\,per~cent of the critical velocity
can have modes up to 6\,d$^{-1}$ \citep{2013MNRAS.429.2500B}. Nevertheless,
while the choice of cutoff frequency is rather arbitrary, this study is
concentrating on slow to moderately rotating stars, where
\citet{2013MNRAS.429.2500B} predict a cutoff closer to 4\,d$^{-1}$. An arbitrary
upper amplitude limit for both $\delta$~Sct and $\gamma$~Dor stars was chosen,
again by inspection of the WASP results, to be 60\,mmag. Fig.~\ref{freq-amp}
suggests that the location of Am pulsators does not appear to be different to
that of the other A stars. There are several high-amplitude, low-frequency
pulsators found which show the characteristics of RR~Lyr stars (labelled `RRL' in
Table~\ref{Table_Results}), plus a small group of higher amplitude
$\delta$~Sct stars (See Sect.~\ref{HighAmp}).

To investigate the influence of atmospheric parameters on the pulsations, we
further selected only those stars for which {\teff} and {\logg} had been
determined by LAMOST. Approximately 90\,per~cent of the original sample have
stellar parameters, which reduced the number in the final sample to 9219, with
808 of them being Am stars. The uncertainties in {\teff} and {\logg} have
average values of $\pm$135\,K and $\pm$0.43\,dex, respectively. The quoted
uncertainty in $\log g$ could possibly be overestimated as it is much larger
than the scatter in the values in the current sample. In addition, the quoted
uncertainties could be larger due to the presence of binary systems. This will mostly affect systems with moderately dissimilar components
and modest brightness ratios, since for nearly equal stars and those with large
brightness ratios the effect will be minimal.

\begin{figure}
\includegraphics[width=\columnwidth]{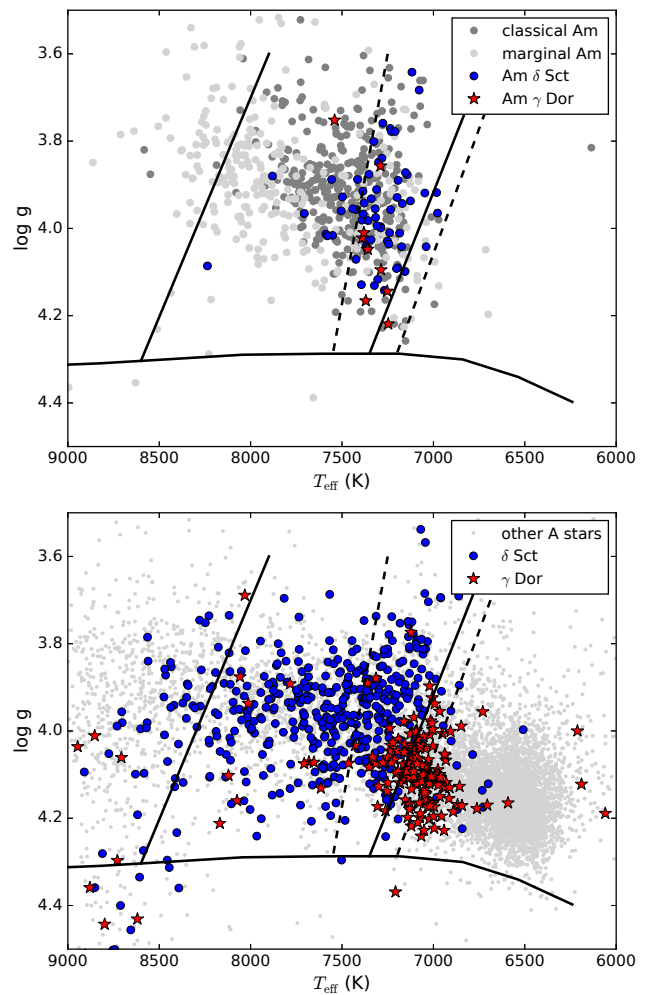}

\caption{Temperature--gravity diagrams showing the location of Am stars (top
panel) and other A stars (bottom panel) in the LAMOST-WASP sample. The location
of the identified $\delta$~Sct and $\gamma$~Dor pulsators is also shown. The
locations of the ground-based $\delta$\,Sct \citep[solid
line;][]{2001A&A...366..178R} and $\gamma$\,Dor \citep[dotted
line;][]{2002MNRAS.333..251H} instability strips are shown, together with the
ZAMS (near-horizontal solid line).}
\label{teff-logg}
\end{figure}

The location of Am and the other A stars in the {\teff}--{\logg} plane is shown
in Fig.~\ref{teff-logg}. The pulsating Am stars are mostly located within the
temperature range $6900 < {\teff} < 7600$\,K, which is toward the red edge of
the $\delta$~Sct instability strip. Note that this region is also dominated by
the classical Am stars. At higher temperatures the decline in pulsating Am
stars appears to occur around the blue edge of the $\gamma$~Dor instability
strip, but this could be coincidence. In contrast, the other A star population
of $\delta$~Sct pulsators extends to much hotter temperatures.

\subsection{$\gamma$ Dor Pulsations}

There are relatively few $\gamma$~Dor Am pulsators in the WASP sample. With the
exception of one possibly misclassified star, they are all to be found within
the ground-based $\gamma$~Dor instability strip. This region also contains the
majority of the $\delta$~Sct Am stars, which suggests that they are probably
hybrids (see Sect~\ref{hybrids}). However, it is not clear whether the low
numbers are related purely to temperature, or if there is interplay between
peculiarity and g-mode pulsation. The incidence of Am stars exhibiting
$\gamma$~Dor pulsations is roughly half that of the other A stars.
However, the lack of many $\gamma$~Dor Am pulsators is consistent with the
relative low number of Am stars in the region where pure g-mode $\gamma$~Dor
pulsators predominately occur. The convective envelopes of $\gamma$~Dor stars
are, on average, deeper than those of $\delta$~Sct stars and mixing by
convective motions causes the peculiarities to vanish. The $\gamma$~Dor
pulsators have frequencies in the range where WASP data are most significantly
affected by the diurnal sampling. Therefore, we will not consider these stars
any further in the current work.

\subsection{Higher amplitude pulsations}
\label{HighAmp}

As mentioned earlier, we identified a small group of $\delta$~Sct stars with
amplitudes $>60$\,mmag. In the temperature range $6900 < {\teff} < 7600$\,K
there are only 10 of these stars, but none of them have been classified as Am.
Randomly picking 10 stars from the WASP sample, using the hypergeometric
probability distribution \citep{2001probstats}, predicts an average of
1.8\,$\pm$\,1.2 Am stars being selected and the probability of not picking any
Am stars is 14\,per~cent. It is, therefore, possible that the lack of Am
stars in this small group could be due to a sampling effect. Indeed, the lack of
any such stars in the \citet{2011A&A...529A..89D} results presented in the
bottom panel of Fig.~\ref{freq-amp} shows that these are not common among the
$\delta$~Sct stars in general. However, we cannot exclude the possibility that
there are no Am stars in this region due to physical mechanisms preventing
either the peculiarities or the oscillations from occurring.

Most RR~Lyr stars have $\teff < 6900$\,K and are therefore too cool to be
considered in the discussion of Am stars. In the WASP sample of RR~Lyr stars
there are only four hotter than 6900\,K. There is one Am candidate found in
the RR~Lyr domain, AL~CMi, but it has no LAMOST stellar parameters.
Literature broadband photometry of this star suggests that $\teff \simeq 7000$\,K,
putting this star within the realm of the Am stars. \citet{1994AJ....108.1016L}
determined a metallicity of $[M/H]$ = $-$0.85 for AL~CMi. However, this was
determined from the relative strengths of the Ca~K and the Balmer lines and
indicates that the calcium line is significantly weaker than expected. The weak
Ca~K line has resulted in \citet{2015MNRAS.449.1401H} misclassifying this RR~Lyr
star as a classical Am star.

\subsection{Binarity}

Brightness variations consistent with being binary systems were found for 249
stars in the WASP sample. The majority appear to show ellipsoidal variations
with periods $\la3$\,d, while a few have Algol-like eclipses. Of the 249 binary
stars, 26 also exhibit $\delta$~Sct pulsations, but none is an Am star. However,
there are only 11 Am stars among the 249 binary systems. Previously, in their
analysis of 1742 Am stars, \citet{2014A&A...564A..69S} found that only 4 out of
70 ($\sim 6$\,per~cent) eclipsing binary systems found also had a pulsating
component detectable with WASP. That pulsations have not been detected in the 11
Am stars in the current WASP sample is not statistically significant. In the
present work we have not performed an in-depth search for eclipsing binary
systems in the WASP light curves and, therefore, will not consider this any
further.

\section{$\delta$ Sct Pulsations}

In this study, we concentrate on the stars with $\delta$~Sct pulsations. These
are selected as having their principal pulsation frequency $>4$\,d$^{-1}$ and an
amplitude $<60$\,mmag. Inspection of Fig.~\ref{freq-amp} reveals that there is a
lack of Am stars with principal frequencies above 40\,d$^{-1}$, while the
\citet{2011A&A...535A...3S} results do contain some Am stars with higher
frequencies. However, this is a selection effect, as the higher frequency
pulsations tend to be found in hotter stars \citep{2000A&AS..144..469R}. All
stars in the WASP sample with principal frequency $>40$\,d$^{-1}$ have {\teff}
$\ga$ 7800\,K, which is hotter than where most of the classical Am stars are
located. Furthermore, in \citet{2011A&A...535A...3S} the high frequency
pulsators all appear to be hotter stars. This was also the case in
\citet{2014MNRAS.439.2078H}, where 12 of 13 high-frequency pulsating Am stars
had ${\teff}>7600$\,K, with 11 having ${\teff}\ge 7800$\,K.

The location of the majority of the Am star pulsators is restricted to the
temperature range $6900 < {\teff} < 7600$\,K. The $\delta$~Sct pulsators among
the other A stars show a wider distribution of effective temperatures,
especially to hotter temperatures. Clearly, there is a difference between the
temperature distributions of pulsating Am stars and pulsating other A stars.
Splitting the Am stars into classical and marginal, using the $\Delta$ index,
reveals that the classical Am stars are concentrated towards the red edge of the
$\delta$~Sct instability strip, while the marginal Am stars are more spread and
peak at slightly hotter temperatures (Fig.~\ref{teff-Am-dist}). The $\Delta$
index is also sensitive to temperature, since the Ca~K line lies on the flat
part of the curve of growth \citep{1978ApJ...221..869K}. Therefore, the marginal
Am stars also contain a contribution from the hot Am stars that may have high
abundance anomalies, but these do not show as strongly in spectroscopic
classification criteria at higher temperatures. In the temperature range $6900 <
{\teff} < 7600$\,K around 13\,per~cent of the classical Am stars have been found
to pulsate, while for the marginal Am stars this rises to around 24\,per~cent.

\subsection{Variation with metallicism}

\begin{figure}
\includegraphics[width=\columnwidth]{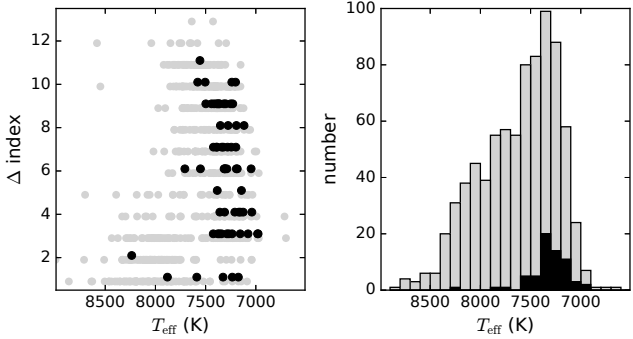}
\caption{The location of Am stars within the {\teff}--$\Delta$ plane (left panel).
The light grey filled circles are the Am stars and the black filled circles are
those found to exhibit $\delta$~Sct pulsations. Note that the different symbols
have been offset slightly in $\Delta$ for clarity. Irrespective of metallicism the
Am pulsators are mostly confined to the region with {\teff}
cooler than $\sim 7600$\,K. The right panel shows the
incidence of Am stars as a function of {\teff} (light grey histogram), along with
those found to pulsate (black histogram). The histogram bin width is 100\,K.}
\label{teff-delta}
\end{figure}

The location of Am pulsators is almost exclusively confined to the region with
{\teff} cooler than $\sim 7600$\,K (Fig.~\ref{teff-delta}). With very few
exceptions, there is no significant variation in the {\teff} range with
metallicism. Hence, the location of Am pulsators appears to be strongly
determined by {\teff} and not significantly dependent on metallicism.

\begin{figure}
\includegraphics[width=\columnwidth]{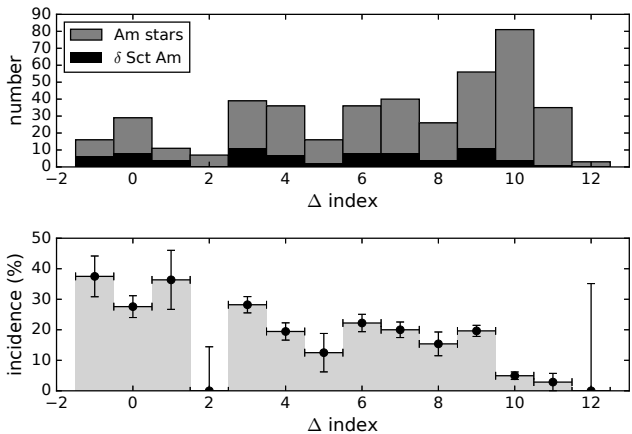}
\caption{The variation of pulsation in Am stars as a function
of metallicism index ($\Delta$). The top panel shows the number of Am stars per
$\Delta$ index bin (grey bars) and the number exhibiting
$\delta$~Sct pulsations (black bars). The lower panel shows the
incidence of the Am pulsations. The errors bars indicate the width of the
$\Delta$ index bin and the uncertainty in the incidence caused by the discrete
nature of the small numbers per bin. In both panels the temperature range is
restricted to $6900 < {\teff} < 7600$\,K.}
\label{Am-delta}
\end{figure}

There are 76 Am stars identified as having $\delta$\,Sct pulsations out of a
total of 439 Am stars within the temperature range $6900 < {\teff} < 7600$\,K, giving an overall
incidence for pulsating Am stars of 17\,per~cent. Fig.~\ref{Am-delta} shows the
variation in the incidence of pulsating Am stars versus $\Delta$ index. There is
a distinct trend with $\Delta$ index, with the incidence of pulsations
decreasing as the metallicism increases. Both Pearson ($-0.71$) and Spearman
($-0.67$) correlation coefficients indicate strong negative correlations, with
confidences $>99$\,per~cent. We, therefore, conclude that classical Am stars are
less likely to exhibit $\delta$\,Sct pulsations than marginal Am stars at the
precision of the WASP photometry.

\subsection{Lower amplitudes in Am stars?}
\label{lower_amp}

\begin{figure}
\includegraphics[width=\columnwidth]{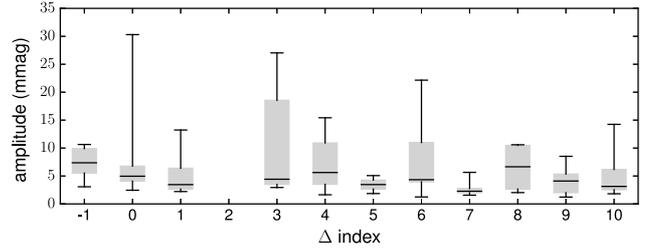}
\caption{The variation of principal pulsation amplitude as a function of
metallicism index ($\Delta$). The boxes
extend from the lower to upper quartile amplitude, with the horizontal lines indicating the median, and the whiskers extending from
the box give the full amplitude range. The temperature range is
restricted to $6900 < {\teff} < 7600$\,K.}
\label{Am-k-amp}
\end{figure}

The results of previous investigations into Am pulsations have led to a
suspicion that they have lower amplitudes than normal $\delta$ Sct stars
\citep{2011A&A...535A...3S}. Fig.~\ref{Am-k-amp} shows that there is no obvious
trend in principal pulsation amplitude with $\Delta$. There does not
appear to be any significant difference in the pulsation amplitudes of classical
Am stars compared to those of the marginal Am stars. The maximum amplitude for
pulsating Am stars is also not significantly different from that of the other A
stars.  Hence, at the milli-magnitude precision of the WASP data, there is no
evidence to support the notion that Am stars might have lower pulsation
amplitudes than other $\delta$~Sct stars.

In their analysis of 29 Am stars observed by {\it Kepler} and {\it K2} Campaign
0, \citet{2015MNRAS.448.1378B} identified 12 stars with $\delta$ Sct pulsations.
Of these, 5 have at least one pulsation amplitude above 1\,mmag and, therefore,
potentially detectable by WASP. The incidence rate (17\,per~cent) is consistent
with that found in the current study. The remaining 7 stars, however, have
amplitudes too low to be detectable by WASP. The distribution of the pulsation
amplitudes for the general population of $\delta$~Sct stars identified by
\citet{2011A&A...529A..89D} has $\sim 60$\,per~cent with amplitudes above 1\,mmag.
Hence, this percentage implies that the probability of selecting 7 or more stars
out of 12 with amplitudes $<1$\,mmag is 16\,per~cent. Therefore, this small
sample does not provide strong significant evidence to support the hypothesis
that pulsation amplitudes of Am stars might be systematically lower than those
of the general population of $\delta$~Sct stars. 

To investigate this further, we have extended the sample of Am stars with {\it
Kepler/K2} photometry as follows. The \citet{2015MNRAS.449.1401H} list of Am
stars includes 67 with {\it Kepler} photometry. These are principally from the
LAMOST-{\it Kepler} project \citep{2015ApJS..220...19D}. The {\it Kepler}
Presearch Data Conditioning \citep[PDC,][]{2010ApJ...713L..87J} light curves of
these, together with the \citet{2015MNRAS.448.1378B} stars, were analysed with {\sc
period04} to obtain their principal pulsation frequency and amplitude.

In addition, {\it K2} has observed further Am stars from the catalogue of
\citet{2009A&A...498..961R} during Campaigns 1 to
4\footnote{{\it K2} Guest Observer programs: GO1014, GO2012, GO3012, GO4045; all P.I.
B. Smalley}. Further inspection of the \citet{2015MNRAS.449.1401H} list revealed
that 16 and 5 Am stars were also observed during Campaigns 0 and 4, respectively.
The {\it K2} light curves, extracted using the $K2P^2$ pipeline
\citep{2015ApJ...806...30L}\footnote{Available from the {\it Kepler} Asteroseismic
Science Operations Centre; \url{http://kasoc.phys.au.dk}}, were also examined using
{\sc period04}.

In total, 144 Am stars were observed by {\it Kepler} and {\it K2}. Of these, 42
($\sim 30$\,per~cent) have principal pulsation frequencies and amplitudes within
the $\delta$~Sct range and 29 stars have amplitudes above 1\,mmag. The fraction
of pulsating Am stars with amplitudes greater than 1\,mmag is 64\,per~cent, which is
consistent with that obtained above from the \citet{2011A&A...529A..89D} sample
of $\delta$~Sct stars. Furthermore, approximately 20\,per~cent of the Am stars have
pulsations with amplitudes detectable by WASP, which is compatible with the
incidence rate found in the current study. There is, however, a lack of Am
pulsators with amplitudes above 10\,mmag, but this could be due to the
relatively low number of pulsators in the {\it Kepler/K2} sample.

\begin{figure}
\includegraphics[width=\columnwidth]{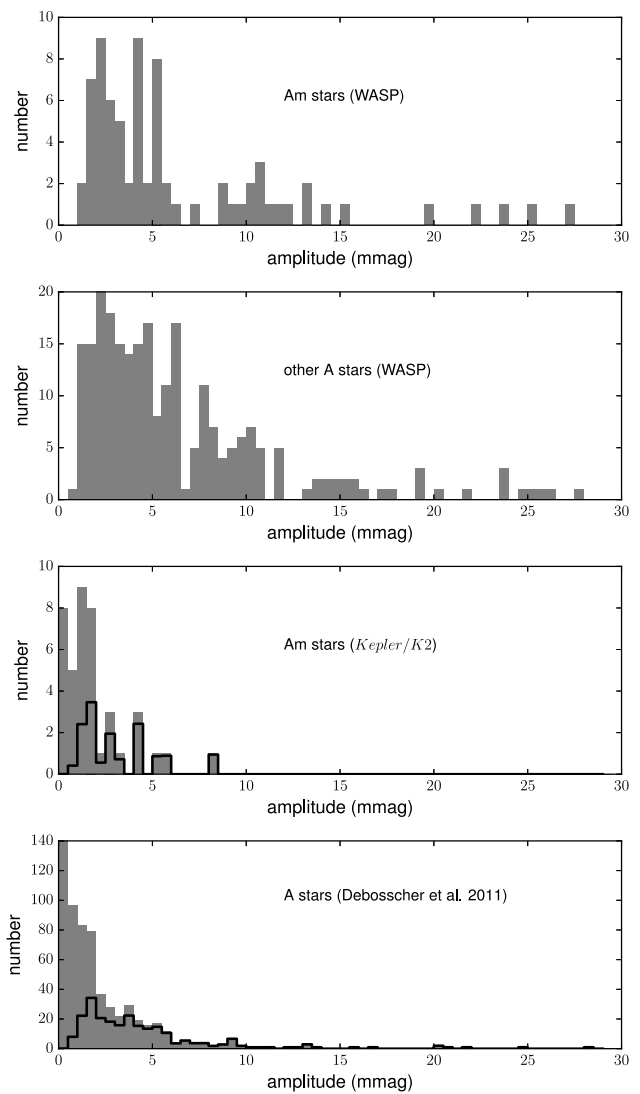}
\caption{The pulsation amplitude distributions for Am and other A stars from
WASP (upper two panels) and {\it Kepler/K2} (lower two panels). These are
presented as solid grey histograms with bin sizes of 0.5\,mmag.
In the lower two panels the outline histogram shows the
amplitude distribution after being convolved with the WASP observability
function to simulate the expected WASP amplitude distribution.
Note the different ordinate scales for the panels.}
\label{Am-amp-dist}
\end{figure}

The amplitude distributions of the Am and other A stars from the WASP and {\it
Kepler/K2} samples are shown in Fig.~\ref{Am-amp-dist}. The amplitude
distributions for the Am and other A stars are broadly similar. In order to
ascertain whether the WASP amplitude distributions are consistent with that
found from the space-based {\it Kepler/K2} photometry, a WASP observability
function was generated from the distribution of the WASP median noise background
measurements (see Fig.~\ref{noise}) and the SNR=4 selection criterion. The
observability function was convolved with the \citet{2011A&A...529A..89D} and
{\it Kepler/K2} amplitude distributions in order to predict the amplitude
distributions for the Am and other A star pulsators expected from the WASP
sample. It is evident from Fig.~\ref{Am-amp-dist} that the observed and expected
WASP distributions for the Am and other A stars are indeed similar.

\subsection{Hybrids}
\label{hybrids}

There are several stars for which both $\delta$~Sct and $\gamma$~Dor frequencies
were detected in the WASP light curves. There are 8 Am and 24 other A stars with
LAMOST stellar parameters in the temperature range $6900 < {\teff} < 7600$\,K. They are
spread around the {\teff}--{\logg} plane within the area occupied by the other
pulsators and not concentrated in any particular region. If these hybrid
pulsators are spread equally among the stars in the above {\teff} range, then
there is a 20\,per~cent probability that at least 8 of them will be Am stars.

In their analysis of early {\it Kepler} data, \citet{2010ApJ...713L.192G}
suggested that hybrid pulsators might be more common among the Am stars.
However, a later study by \citet{2015MNRAS.452.3073B} did not support this
suggestion, finding that at {\it Kepler} precision hybrid pulsators are very
common. Of the 8 Am stars with hybrid pulsations, 5 are classical and 3 are
marginal. Picking 8 stars at random from the 387 Am stars in the temperature
range $6900 < {\teff} < 7600$\,K results in a 74\,per~cent probability of at
least 5 classical Am stars being selected. However, the number of hybrids found
in the present work is rather too low to provide meaningful statistics.

\section{Pulsation Modelling}

In their analysis of the pulsating Am star HD\,187547,
\citet{2014ApJ...796..118A} suggested that turbulent pressure is responsible for
a large part of the excitation in the H/He\,{\sc i} ionisation layer. Using
non-adiabatic pulsational stability analyses
\citep{1992MNRAS.255..603B,1999A&A...351..582H} with time-dependent non-local
convection treatment models \citep{1977LNP....71...15G,1977ApJ...214..196G} the
authors found that only the lowest radial orders were excited by the
$\kappa$-mechanism in the He\,{\sc ii} ionisation layer. Surprisingly, most
pulsation modes were driven by the turbulent pressure in the H/He\,{\sc i} layer
and not in the He\,{\sc ii} ionisation layer. Furthermore, oscillations can be
excited by turbulent pressure acting in the H/He\,{\sc i} ionisation layer,
without any contribution from the $\kappa$-mechanism. These findings are
compatible with pulsations in many Am stars, which otherwise remain unexplained.

Using the models described above, we show non-adiabatic pulsational stability
analyses of a star with global stellar parameters representative of Am stars:
$M$=2.3\,M$_{\odot}$, $T_{\rm eff}$=7500\,K, $L$=25.7\,L$_{\odot}$. The
non-local convection parameters $a$, $b$ and $c$ for the theory of
\citet{1977LNP....71...15G} were chosen to be $a^2 = b^2 = c^2 = 950$. We used
the OPAL opacity tables \citep{1995ASPC...78...31R} supplemented by the
\citet{2005ApJ...623..585F} tables at low temperatures. It is beyond of the
scope of the current work to go into details concerning the theoretical aspect
of mode stability, but for details we refer the reader to
\citet{1999A&A...351..582H} and \citet{2014ApJ...796..118A}.

\begin{figure}
\includegraphics[width=\columnwidth]{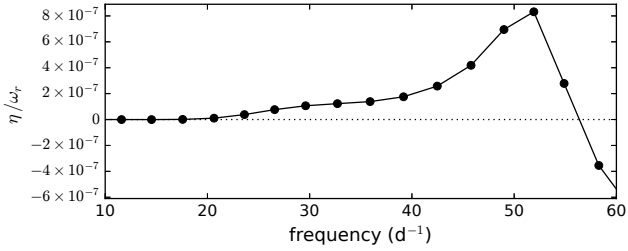}
\caption{Normalized growth rate ($\eta/\omega_r$) as a function of frequency
for the model described in the text. A positive growth rate indicates
excited pulsation modes, while a negative value indicates damped
modes.}
\label{growth}
\end{figure}

\begin{figure}
\includegraphics[width=\columnwidth]{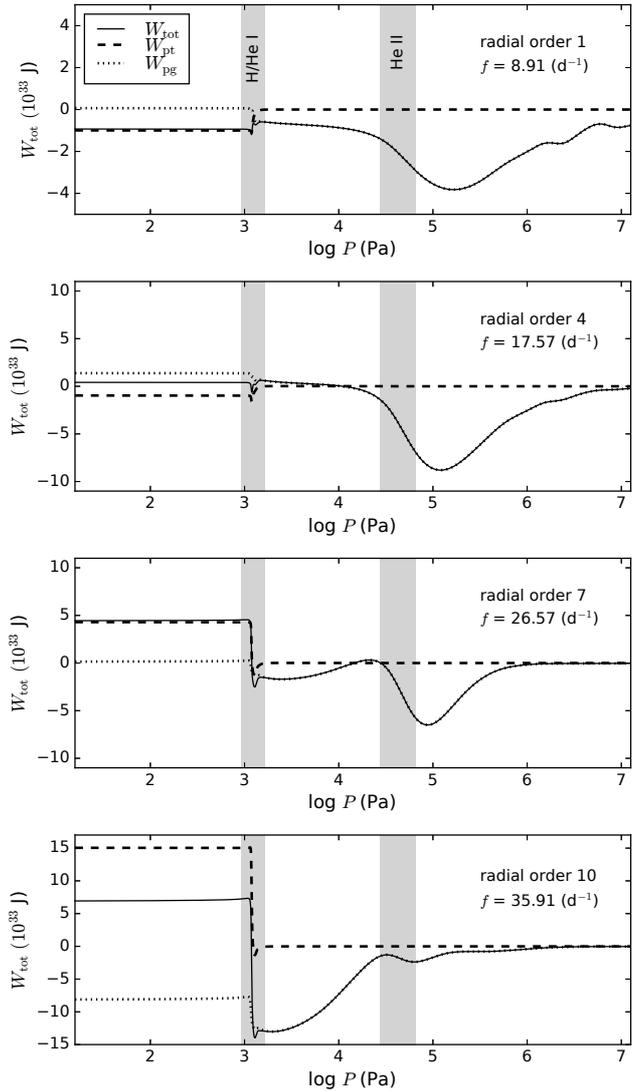}

\caption{The accumulated work integrals ($W_{\rm tot}$) as a function of total
pressure ($\log P$) for four different radial orders. The continuous lines
depict the total accumulated work integral, which when positive at the
stellar surface means that the pulsation mode is intrinsically excited.
As shown in the top panel, the dashed and dotted lines reflect the
contribution of turbulent ($W_{\rm pt}$) and gas ($W_{\rm pg}$) pressure,
respectively, to the total accumulated work integral. As indicated in the top
panel, the grey areas depict the H/He\,{\sc i} and the He\,{\sc ii}
ionisation layers and correspond to the convection zones in a star.}
\label{modes}
\end{figure}

In Fig.~\ref{growth} we show the normalised growth rates ($\eta/\omega_r$, where
$\eta$ and $\omega_r$ are, respectively, the imaginary and real parts of the
eigenfrequency) as a function of frequency. Whenever $\eta/\omega_r > 0 $ the
pulsation mode is intrinsically excited and $\eta/\omega_r < 0$ indicates an
intrinsically damped mode. To further illustrate the driving agent of these
modes we show the accumulated work integrals as a function of total pressure for
four different radial orders (Fig.~\ref{modes}). The continuous lines depict the
total accumulated work integral, which when positive at the stellar surface
means that the mode is intrinsically excited. The dashed and dotted lines
reflect the contribution of turbulent and gas pressure, respectively, to the
total accumulated work integral. The grey areas depict the H/He\,{\sc i} and the
He\,{\sc ii} ionisation layers and correspond to the convection zones in a star
with the parameters outlined above. Note, however, that these zones are likely
linked through overshooting and that additional convection zones may occur
deeper into the star due to atomic diffusion
\citep[e.g.][]{2012A&A...546A.100T,2016A&A...589A.140D}.

As illustrated in Fig.~\ref{modes} the radial fundamental mode (order 1) is
damped because the total accumulated work integral is not positive. There is
some driving due to gas pressure in the He\,{\sc ii} ionisation layer, however,
the mode is damped in the H/He\,{\sc i} layer. The fourth radial order mode, on
the other hand, is intrinsically excited by gas pressure in the He\,{\sc ii}
layer. This is illustrated by the local increase on the accumulated work
integral. Also, in this case, turbulent pressure has a slight damping effect on
the mode. The 7th radial order mode is clearly intrinsically excited by both,
turbulent and gas pressure in the H/He\,{\sc i} and He\,{\sc ii} layers,
respectively. For the 10th radial order mode, however, the excitation
predominantly occurs due to the turbulent pressure in the H/He\,{\sc i}
ionisation layer.

In order to conclude whether turbulent pressure can fully explain the phenomenon
of pulsation in Am stars, a more global study covering the entire instability
strip is required (Antoci et al. in prep.). The current model does not include
the diffusion of He. Adding He depletion increases the range of excited modes,
which will be explored in detail in the forthcoming study. Nevertheless, there
are strong indications that both turbulent pressure and excitation reach a
maximum around 7500\,K \citep{2014ApJ...796..118A,2015A&A...584L...2G}, which is
in the region where pulsating Am stars are found in our study.

\section{Discussion and conclusion}

From a large sample of A and Am stars observed with LAMOST and WASP we have
found that $\delta$~Sct pulsations in Am stars are mostly confined to a region
close to the red edge of the $\delta$~Sct instability strip. Their location is
generally within the temperature range $6900 < {\teff} < 7600$\,K and is
independent of metallicism. The incidence of pulsations in Am stars decreases
noticeably in more peculiar Am stars. The reason for the decrease is not
known, but we might expect the $\kappa$-mechanism and turbulent pressure to
interact depending on several parameters, including the diffusion of helium,
stellar mass and evolutionary stage, as well as the global metallicity.
Nevertheless, the observations presented in the current work are consistent with the hypothesis proposed by
\citet{2014PhDT.......131M} that the non-pulsating stars in the $\delta$~Sct
instability strip are Am stars. The amplitudes of pulsations found with WASP
agree with expectations based on the results from {\it Kepler}. There is some,
albeit rather weak, evidence for lower amplitudes for Am pulsators compared to
other A stars.

The mystery of why Am stars pulsate (or at least some of them) originates from
the assumption that it is the $\kappa$-mechanism operating in the He\,{\sc ii}
ionisation layer that excites $\delta$~Sct oscillations. The blue edge of the
$\delta$~Sct instability strip moves redward as helium abundance is reduced,
while the red edges remain almost constant
\citep{1979ApJ...231..798C,2000A&A...360..603T,2015ChA&A..39...16X}. This is
consistent with the location of pulsating Am stars found in our study. However,
the lack of any significant correlation of location and amplitude with
metallicism suggests that another mechanism may be at play. We would expect
that as metallicism increases and, by inference, helium abundance in the
He\,{\sc ii} ionisation layer decreases, the location of pulsating Am stars
should be further marginalised to cooler temperatures.

In pulsating Am stars we suggest that a large part of the excitation occurring
in the H/He\,{\sc i} ionisation layer is driven by turbulent pressure. Low
radial order modes in normal $\delta$~Sct stars are excited by the
$\kappa$-mechanism. However, the higher ones are predominantly driven by
turbulent pressure in a temperature-dependent manner, since turbulent pressure
damps oscillations in stars close to the red edge of the $\delta$~Sct
instability strip. While the WASP data in this work indicate that Am and
normal $\delta$~Sct stars do not show oscillations with frequencies higher than
around 30\,d$^{-1}$, we know from other WASP and {\it Kepler} studies that there are Am and
non-Am $\delta$~Sct stars showing pulsations up to 80--90\,d$^{-1}$. For
example, HD\,187547 is an Am star with pulsations covering a frequency range from
20--80\,d$^{-1}$ \citep{2014ApJ...796..118A}. The amplitudes of the highest
radial order modes are lower than 0.15\,mmag. WASP could not detect these
amplitudes.  We conclude that it is plausible that Am stars have on average
higher radial order modes excited due to the turbulent pressure being more
efficient as the depleted He is offset by an increase in hydrogen number
density.

\section*{Acknowledgements}

The WASP project is funded and operated by Queen's University Belfast, the
Universities of Keele, St. Andrews and Leicester, the Open University, the Isaac
Newton Group, the Instituto de Astrofisica de Canarias, the South African       
Astronomical Observatory and by the UK Science and Technology Facilities Council
(STFC). Funding for the Stellar Astrophysics Centre was provided by The Danish
National Research Foundation (grant No. DNRF106). The research is supported by
the ASTERISK project (ASTERoseismic Investigations with SONG and Kepler) funded
by the European Research Council (Grant agreement No. 267864).  DWK is supported
by the STFC.  DLH acknowledges support from the STFC via grant number ST/M000877/1.
SJM was supported by the Australian Research Council.

\bsp	
\label{lastpage}
\end{document}